\documentstyle[preprint,aps]{revtex}
\begin{document}
\title{Momentum-dependent mean field based upon the Dirac-Brueckner
approach for nuclear matter}
\author{G. Q. Li and R. Machleidt}
\address{Department of Physics, University of Idaho, Moscow, ID 83843, USA}
\maketitle

\begin{abstract}
A
momentum-dependent mean field potential, suitable for application in the
transport-model description of nucleus-nucleus collisions, is derived
in a microscopic way. The derivation is based upon the Bonn
meson-exchange model for the nucleon-nucleon interaction and
the Dirac-Brueckner approach for nuclear matter.
The properties of the microscopic mean field are examined and compared
with phenomenological parametrizations which are commonly
used in transport-model calculations.
\end{abstract}

The microscopic description of nuclear matter, finite nuclei, and
nuclear reactions in terms of the realistic nucleon-nucleon (NN)
interaction continues to be an interesting topic in nuclear physics.
Although quantum chromodynamics (QCD) is believed to be the ultimate
theory of strong interaction, the only quantitative NN potential models
available up till now are based on the idea of meson exchange;
a well-known example is
the Bonn potential \cite{mach1,mach2}.

Special many-body theories, such as
the Brueckner approach
and the variational method,
have been developed such that a  realistic NN interaction can be applied
in nuclear many-body calculations.
Nuclear matter saturation has been quantitatively explained
by
the Dirac-Brueckner-Hartree-Fock (DBHF) approach \cite{shak,mal,broc1,broc11},
starting from a realistic NN interaction.
Thus, this approach
provides a natural starting point for the self-consistent description
of nuclear matter, nuclear structure, and nuclear reactions in terms of
the realistic NN interaction.

The extension of the DBHF approach from nuclear matter to the
structure of finite nuclei and nucleon-nucleus scattering have
been attempted \cite{may,muth1,broc2,FMM93,li1}.
The direct solution of the DBHF equation
in finite nuclei is, however, rather involved. More practically,
one either parametrizes the DBHF results in nuclear matter in terms
of simple Lagrangians \cite{broc2,FMM93,li1,pol} that can be easily
applied in finite nuclei,
or one performes a full Brueckner-Hartree-Fock (BHF)
 calculation in the finite nucleus
taking the relativistic effects via the local-density approximation
into account~\cite{muth1}.

In tune with the latter
approach,  we present in this paper a momentum-dependent mean field
single-particle potential derived from DBHF nuclear matter calculations.
This mean field potential is suitable for application in
the transport-model description of nucleus-nucleus collisions.
Together with the in-medium NN cross sections derived in
our earlier work \cite{li2}, a fully self-consistent
calculation of nucleus-nucleus collisions can then be performed.

A self-consistent approach has been pursued in Refs.
\cite{aich1,aich2,li4,khoa} applying the Reid soft-core potential~\cite{Rei68}
in a non-relativistic Brueckner-Hartree-Fock
(BHF) calculation. The major criticism
\cite{bert} of
this study comes from the fact that the BHF calculation cannot even
reproduce quantitatively the saturation properties of static nuclear
matter \cite{Day}; the application of its
results to colliding nuclear matter is thus debatable.

 From the theoretical point of view, nucleus-nucleus collisions at
intermediate energies are of particular interest. At these energies,
both the mean field and two-body collisions play an equally important
role in the dynamical evolution of the two colliding systems; therefore,
they have to
be taken into accout on an equal footing in transport models such as
the Boltzmann-Uehling-Uhlenbeck (BUU) equation \cite{bert} or quantum
molecular dynamics (QMD) \cite{aich3}.

In the early applications of transport models to nucleus-nucleus
collisions, the mean field and the in-medium NN cross sections were
parametrized separately: a simple Skyrme-type parametrization was used for the
mean field \cite{bert}, while for the in-medium NN cross sections
free-space
proton-proton ($pp$) data were applied \cite{cugn,bert}. By varying the
mean field (i.e., the parameters in the Skyrme parametrization) ---
which is related to the nuclear equation of state --- and by comparing
theoretical results with experimental data, one expects to obtain some
information on the nuclear equation of state.

There are at least two uncertainties in this approach:
the mean field and the
in-medium NN collisions. Some observables that are believed to carry useful
information about the nuclear equation of state are also strongly affected
by what is assumed for the NN collisions.
Therefore, the in-medium NN cross sections used in transport
models should be  determined consistently before reliable information
concerning the equation of state can be obtained. Within the DBHF approach, the
in-medium NN cross sections below pion threshold have been studied
in Ref. \cite{li2}. Using those microscopic in-medium NN cross sections
will reduce the uncertainty concerning the in-medium NN collisions.

Furthermore, it is well
known that ---
due to correlations and exchange terms ---
the effective nuclear interaction is momentum dependent.
Empirically,
the momentum dependence of the mean field is best illustrated by
the nucleon effective mass $m^*/m$ which is about 0.7 at low energies
and approaches unity at high energies \cite{mah}.
Momentum dependence of the mean field has been found to play an important role
in observables that are claimed to carry useful information about the nuclear
equation of state, such as transverse momentum transfer \cite{aich4,gale}
and particle production
\cite{li5} in nucleus-nucleus collsions.
The  microscopically derived momentum dependence of the mean field
may help reducing  also this source of uncertainty.

Our derivation of the mean-field potential is based upon the DBHF
approach for nuclear matter. For a comprehensive and pedagogical
introduction into the basic ideas of the Dirac-Brueckner approach
and its formalism, we
refer the interested reader
to section 10.5 of Ref.~\cite{mach2} and Ref.~\cite{broc11}.
Here, we will only briefly summarize the major points and the
basic equations.
For the NN potential, we will
apply the one-boson-exchange representation of the
Bonn meson-exchange model for the NN interaction~\cite{BonnA}.
It includes six nonstrange mesons
with given mass and coupling.
Pseudovector (derivative) coupling is used for
pseudoscalar mesons ($\pi $ and $\eta $).
A  form factor of monopole type
is applied at each meson-nucleon vertex which simulates the short-range
dynamics of quark-gluon nature.
The potential provides an accurate description of the deuteron and NN
scattering \cite{mach2,broc11}.

The essential point of the DBHF
approach is the use of the Dirac equation for the description of
single-particle
motion in the nuclear medium
\begin{equation}
[{\bbox \alpha \cdot }{\bf p}+\beta (m+U_S)+U_V]\tilde u ({\bf p},s)=
\epsilon\tilde u ({\bf p},s)
\end{equation}
where  $U_S$ is an attractive scalar field and $U_V$
the time-like component of
a vector field which is repulsive;
$m$ denotes the experimental mass of the free nucleon.

The positive-energy Dirac spinors that solve Eq.~(1)
can be written as
\begin{equation}
\tilde u ({\bf p},s) = \left( \frac{\tilde E_{\bf p} +\tilde m}
{2\tilde m} \right)^{\frac12} \left( \begin{array}{c}
                                          1   \\
       \frac{{\bbox \sigma} \cdot {\bf p}}
             {\tilde E_{\bf p} + \tilde m}
                                     \end{array} \right)
  \chi_s
\end{equation}
with
\begin{eqnarray}
\tilde m & = & m + U_S \; , \\
\tilde E_{\bf p} & = & (\tilde m ^2 + {\bf p}^2 )^{\frac12}
\end{eqnarray}
and $\chi_s$ a Pauli spinor.
The normalization is $\bar{\tilde u} u = 1$.

Similar to conventional Brueckner theory, the basic quantity in the
DBHF  approach is a $\tilde G$-matrix which satisfies
the in-medium Thompson equation (relativistic Bethe-Goldstone equation)
\cite{mach2,broc11,li6}
\begin{eqnarray}
\tilde G({\bf q',q; P})&=&\tilde V({\bf q',q})
+{\cal P}\int {d^3k\over (2\pi )^3}\tilde V({\bf q',k}){\tilde m^2\over
\tilde E^2_{(1/2){\bf P+k}}} \nonumber \\
 & & \times {Q({\bf k,P})\over
2\tilde E_{(1/2){\bf P+q}}-2\tilde E_{(1/2){\bf P+k}} }
\tilde G({\bf k,q;P})
\end{eqnarray}
For a pair of interacting nucleons with momentum ${\bf p}_1$ and ${\bf p}_2$
in the nuclear matter rest frame,
the center-of-mass (c.m.) momentum is ${\bf P}={\bf p}_1+{\bf
p}_2$ and their relative momentum ${\bf q}=(1/2)({\bf p}_1-{\bf p}_2)$.
$\cal P$ denotes the principal value.
Note that Eq.~(5) is density dependent due to the Pauli projector, $Q$,
and the scalar field, $U_S$; in the notation used in Eq.~(5),
this density dependence is suppressed.
Notice also that $\tilde{V}$ is density dependent due to the in-medium Dirac
spinors, Eq.~(2), representing the four outer legs of the one-meson-exchange
Feynman diagrams defining the potential. The in-medium Dirac
spinors depend on $\tilde{m}$ which through $U_S$ depends on the
density of the medium.

The Dirac equation and the in-medium Thompson equation are solved
self-consistently. The nuclear matter properties are derived from
the $\tilde G$-matrix.
For more details, see section 10.5 of Ref.~\cite{mach2} and
Ref.~\cite{broc11}.

According to Eq~(1),
the single-particle energy
of a nucleon with momentum ${\bf p}_i$ is
\begin{eqnarray}
\epsilon_i & = &
{\tilde m\over \tilde E_i}<i|{\bbox \gamma \cdot}{\bf p}_i+m|i>
+{\tilde m\over \tilde E_i}U_S + U_V
\\
 & = &
{m\tilde m+{\bf p}_i^2\over \tilde E_i}
+{\tilde m\over \tilde E_i}U_S + U_V
\\
 & = & \tilde E_i+U_V
\end{eqnarray}
In our notation, $|i>$ is a single particle state represented
by a Dirac spinor of the kind Eq.~(2)
and $<i|$ is the {\it adjoint} Dirac spinor,
$\bar{\tilde u}\equiv \tilde u ^\dagger \gamma_o$;
$<i|i>=1$;
$\beta \mbox{\boldmath $\gamma$} = \mbox{\boldmath $\alpha$}$;
$\beta = \gamma_0$.

The scalar and vector field, $U_S$ and $U_V$, are determined from
$${\tilde m\over \tilde E_i}U_S+U_V={\rm Re} \sum _{j\le k_F}
{\tilde m\over \tilde E_i\tilde E_j}<ij|\tilde G|ij-ji> \eqno (9)$$
This is the
relativistic analogue to the non-relativistic Brueckner-Hartree-Fock
definition of a single-particle potential using the effective
mass approximation.
Self-consistency of the DBHF calculations
means satisfying Eq.~(6).

As shown in Ref.~\cite{mach3},
it is a good approximation to assume
$U_S$ and $U_V$ to be momentum independent
(note, however, that they are density dependent).
Notice also that momentum-independence of $U_S$ and $U_V$
does {\it not} imply momentum-indepence of the mean field (cf.\
Eq.~(13), below).
Momentum-dependence of $U_S$ and $U_V$
is one possible source of momentum-dependence of the
mean field, but it is the less important one
(see discussion below).

In order to define a mean-field potential,
we rearrange the single-particle energy, Eq.~(8),
$${\epsilon_i=E_i+U_V+\tilde E_i-E_i=E_i+U_i}\eqno (10)$$
with the free energy
$${E_i=({\bf p}_i^2+m^2)^{1/2} \: ,}\eqno (11)$$
and the mean-field potential
$${U_i=U_V+\tilde E_i-E_i \; .}\eqno (12) $$
More explicitly, this mean field is
$${U(\rho,p)=U_V(\rho)+[(m+U_S(\rho))^2+{\bf p}^2]^{1/2}
-(m^2+{\bf p}^2)^{1/2}
}\eqno (13)$$
where the dependence on the density $\rho$ is now clearly indicated
($p\equiv |{\bf p}|$).
$U_S(\rho)$ and $U_V(\rho)$ are listed for many densities
in Table VII of Ref.~\cite{broc11}; e.~g., for $\rho_0=0.17$ fm$^{-3}$
one has $U_S=-355.7$ MeV and $U_V=274.7$ MeV. By using this table in a
spline interpolation (adding $U_S=U_V=0$ for $\rho=0$), one can obtain
$U_S$ and $U_V$ with good accuracy for any density between 0 and
$4.2 \rho_0$. For a simple parametrization of our mean field potential,
see Eq.~(20), below.

The mean field, Eq.~(13), shows strong momentum dependence
for small $p$. For large $p$ ($p>>m$), the square roots are $\approx p$
and, thus, they cancel. Therefore, $U$ vanishes approximately
for large $p$ (notice that $U_V$ can be neglected
when $p$ becomes very large). In summary,
$U$ behaves in a physically reasonable way.

As mentioned, we have neglected the momentum
dependence of $U_S$ and $U_V$. In Ref.~\cite{mach3} it is shown that,
for low momenta, the momentum dependence of $U_S$ and $U_V$ can be
described by
\setcounter{equation}{13}
\begin{eqnarray}
U_S(p) & = & U_S^{(0)} - U_S^{(1)} \frac{p^2}{p^2_F} \\
U_V(p) & = & U_V^{(0)} - U_V^{(1)} \frac{p^2}{p^2_F}
\end{eqnarray}
with $p_F$ the Fermi momentum.
It is found that~\cite{shak,mach3}
\begin{equation}
\frac{U_S^{(1)}}{U_S^{(0)}}\approx
\frac{U_V^{(1)}}{U_V^{(0)}}\approx
0.05
\end{equation}
Expanding Eq.~(13) up to terms in $p^2$ and using Eqs.~(14)-(16),
one obtains
\begin{equation}
U(p)  \approx  U_V^{(0)} + U_S^{(0)}
-0.05 \frac{U_V^{(0)}+U_S^{(0)}}{p^2_F} p^2
+ \frac{m-\tilde m}{2m\tilde m} p^2
\end{equation}
The last term on the r.h.s. describes the momentum dependence of Eq.~(13)
for low momenta; the second but last term (which is typically a factor of five
smaller) is the additional momentum dependence from $U_S$ and
$U_V$ that is obviously negligible. Notice also that $U_S$ and $U_V$
are of opposite sign, but of the same order of magnitude.

The nucleon effective mass, $m^*$, as predicted by the DBHF calculations
for nuclear matter is discussed in detail in Ref.~\cite{li6}.
Using the Bonn~A potential, which we apply in this work, one gets
$m^*/m=0.71$ at normal nuclear matter density ($\rho_0=0.17$ fm$^{-3}$)
--- a reasonable value.

Empirically, the most sophisticated investigation of the nucleon mean field
(optical potential) is conducted in Dirac phenomenology
\cite{clar,coop}. In this approach, the
scalar and vector potentials are parametrized with a number of free
parameters fitted to nucleon-nucleus scattering data. In order to compare
our microscopic mean field obtained in DBHF calculations with the
empirical one from Dirac phenomenology, we extract from the Dirac global
code \cite{coop} the scalar and vector potentials in the centers of
$^{40}$Ca and  $^{208}$
Pb.  In Fig. 1 we
compare our microscopic mean field (solid curve)
at normal nuclear matter ($\rho _0$=0.17
fm$^{-3}$) with those extracted from the Dirac phenomenology. The open circles
and squares correspond to empirical values in the center of
$^{208}$Pb based on  fit 1 and fit 3   of
Ref. \cite{coop}, respectively, while the solid triangles
are the empirical values in the center of $^{40}$Ca based on fit 1
of Ref.~\cite{coop}. The difference between the circles and squares
reflects the uncertainties in the empirical determination of the
mean field, while the difference between the circles and triangles
shows, to some extent, finite-size effects.
In general, our microscopic result is
in good agreement with the empirical one. The microscopic mean field,
shows, however, a stronger momentum dependence. This may be due to
the assumption of momentum independence of the scalar and vector
fields in our DBHF calculation. Finite-size and surface effects,
included in the Dirac phenomenology, could also play a role.

In the past, phenomenological parametrizations of momentum-dependent
mean-field potentials have been constructed and used in transport models.
We mention here two typical examples. One is proposed in Ref. \cite{gale}
and used in BUU calculations \cite{bert}, we denote it by $U^{(1)}$;
the other is proposed in Ref. \cite{aich4} and used in QMD calculations
\cite{aich3}, we denote it by $U^{(2)}$. For $U^{(1)}$, one parameter
set has been given \cite{bert,gale} that corresponds to a soft equation
of state with incompressibility $K=$ 215 MeV; we denote this parameter set by
GBG. For $U^{(2)}$, two parameter sets, corresponding to a
soft and hard equation of state, have been proposed \cite{aich3,aich4}; they
are usually denoted by SMD ($K$=200 MeV) and HMD ($K$=380 MeV),
respectively.
In static nuclear matter, $U^{(1)}$ and $U^{(2)}$ are given by
$${U^{(1)}(\rho , p)=\alpha \rho +\beta \rho ^\gamma
+\delta(0.7965+
{1\over 1+{{\bf p}^2/ \Lambda ^2}})\rho }\eqno (18)$$
$${U^{(2)}(\rho , p)=\alpha \rho +\beta \rho ^\gamma
+\delta {\rm ln}^2(\varepsilon {\bf p}^2+1)\rho ^\sigma}
\eqno (19)$$
Parameter sets for $U^{(1)}$ (GBG)
and $U^{(2)}$ (SMD, HMD) are listed in Table~1.

In Fig. 2, we show the density dependence
of our microscopic (Eq. (13)) and of phenomenological (Eqs. (18) and (19))
mean fields. At low densities, the four mean fields shown
are very close to each other. At high densities, there are significant
differences: the HMD mean  field is much stiffer than the others; the
microscopic mean field is more close to the soft (SMD and GBG) phenomenological
mean fields. There is also an important difference between the microscopic
mean field obtained in our (relativistic)
DBHF calculation and the one from non-relativistic BHF:
the former is stiffer than the latter (see Fig. 4 of Ref. \cite{aich2}
and Fig. 3 of Ref. \cite{khoa}) due to repulsive relativistic effects.

In Fig. 3, we show the momentum dependence of the microscopic and
phenomenological mean fields. Both microscopic and phenomenological
mean fields increase with increasing momenta.
For the momenta shown, the
microscopic mean field has a stronger momentum dependence than the
phenomenological ones. Such a strong momentum dependence is also
observed in the microscopic mean field from non-relativistic
BHF calculations \cite{khoa}.
Note, however, that for very large momenta (which are not shown in Fig.~3)
the situation is reversed: the momentum-dependence disappears from our
microscopic mean field (as discussed above), while the phenomenological
parametrization continues to grow.

For convenience in applications, we have parametrized
our microscopically derived mean field using exactly the same
{\it ansatz} as for $U^{(2)}$, Eq.~(19), namely,
$${U(\rho ,p)=\alpha \rho +\beta \rho ^\gamma+
\delta {\rm ln}^2(\varepsilon {\bf p}^2+1)\rho ^\sigma}
\eqno (20)$$
The parameters are listed in Table 1, row `DBHF'.
The quality of the fit can be seen in Fig. 4 where
solid curves are obtained from Eq. (20) while the circles
correspond to the exact calculation using Eq.~(13).
 Both density dependence (upper part of Fig.~4)
and momentum dependence (lower part of Fig.~4) are
reproduced well by the parametrization Eq. (20), for the densities
and momenta shown.
Note, however, that the parametrization Eq.~(20) is bad for large $p$
($p>>m$) since it continues to grow with increasing $p$,
while the exact mean field, Eq.~(13), becomes independent
of $p$ for large momenta.
These very large momenta may, however, not be important in
nucleus-nucleus collisions at intermediate energies.

The incorporation of the microscopic mean field and the
in-medium NN cross sections
in transport-model calculations is straightforward.
The propagation
of nucleons is  determined by Hamilton's equations of motion
\cite{bert,aich3}
$${{d{\bf p}\over dt}=-\nabla _{\bf r}U}\eqno (21a)$$
$${{d{\bf r}\over dt}={{\bf p}\over ({\bf p}^2+m^2)^{1/2}}+\nabla _{\bf p}
U}\eqno (21b)$$
where the position dependence of the mean field enters through the
position dependence of the density via the local-density approximation.
Moreover,
using microscopic in-medium NN cross sections \cite{li2},
the in-medium collisions
can be treated in the same way as in usual transport models.
The self-consistent
investigation of nucleus-nucleus collisions based upon realisitic NN
interactions will be reported elsewhere.

In summary, we have derived a microscopic momentum
dependent mean field using the
DBHF approach for nuclear matter and the Bonn potential.
The momentum dependence of our mean field comes out very close to that
obtained in Dirac phenomenology.
Qualitatively, our microscopic mean field
shows similar density and momentum dependence as some of the phenomenological
parametrizations, which are
 often used in BUU and QMD calculations; quantitatively, however, there
are important differences, especially at high densities and momenta.
For practical purposes, we have parametrized our
microscopic mean field in terms of a simple analytic
form which accurately reproduces our exact results; this can easily be
applied in transport-model calculations.

\vskip 1cm
{\bf Acknowledgement}.
This work is supported in part by the U.S. National
Science Foundation under Grant No. PHY-9211607 and by the Idaho
State Board of Education.
\pagebreak

\pagebreak
{\bf Table 1}: Parameters of phenomenological and microscopic
mean fields.
The parameter set `DBHF'
is the parametrization
of our microscopic mean field, Eq.~(13), in terms of the {\it ansatz} Eq. (20).
GBG uses Eq.~(18), while SMD and HMD use Eq.~(19).
\vskip 0.3cm
\begin{center}
\begin{tabular}{cccccccc}
\hline\hline
 &$\alpha$ (MeV$\cdot$fm$^{3}$)  &$\beta$ (MeV$\cdot$fm$^{3\gamma}$) &
$\gamma$ & $\delta$ (MeV$\cdot$fm$^{3\sigma}$)
&$\varepsilon$ (MeV$^{-2}$)& $\sigma$ & $\Lambda$ (MeV) \\
\hline
 GBG & --888.96 & 1687.53 & 7/6 & --460.12 & --- & --- & 400 \\
 SMD & --2294.12& 2412.34 & 1.14& 9.24 & 5.0$\times$10$^{-4}$ & 1.0 & ---\\
 HMD & --764.71 & 2394.49 & 2.09& 9.24 & 5.0$\times$10$^{-4}$ & 1.0 & ---\\
 DBHF& --1129.51& 1501.31 & 1.48& 35.07& 4.3$\times$10$^{-5}$ & 0.7 & ---\\
\hline\hline
\end{tabular}
\end{center}

\pagebreak

\begin{figure}
\caption{Momentum dependence of empirical and microscopically derived
mean field potentials.
The microscopic mean field is calculated for nuclear matter with $\rho =
0.17$ fm$^{-3}$. The empirical mean fields are extracted from
Dirac phenomenology at the centers of $^{40}$Ca and $^{208}$Pb, the
open circles
and squares correspond to $^{208}$Pb based on
fit 1 and fit 3 of Ref.~[36],
while the solid triangles correspond to $^{40}$Ca based on
fit 1.}
\end{figure}

\begin{figure}
\caption{Density dependence of microscopic and phenomenological
mean fields for several nucleon momenta.
The solid curve represents our microscopic mean field
as obtained from DBHF. The dashed and dash-dotted lines are
phenomenological parametrizations proposed in Ref.~[24].
The dotted line is from Ref.~[19,25].}
\end{figure}

\begin{figure}
\caption{Momentum dependence of microscopic and phenomenological
mean fields for several densities.
Notation as in Fig.~2.}
\end{figure}

\begin{figure}
\caption{Comparison of the parametrization Eq.~(20) (solid curves) with the
exact results (dots) for our microscopic mean field, Eq.~(13).
The upper part shows the density dependence, and the
lower part the momentum dependence.}
\end{figure}


\begin{thebibliography}{99}
\bibitem{BonnA} In this paper, we apply the so-called Bonn-A potential
which is defined in Table A.2
of Ref.~\cite{mach2}.
\end{thebibliography}
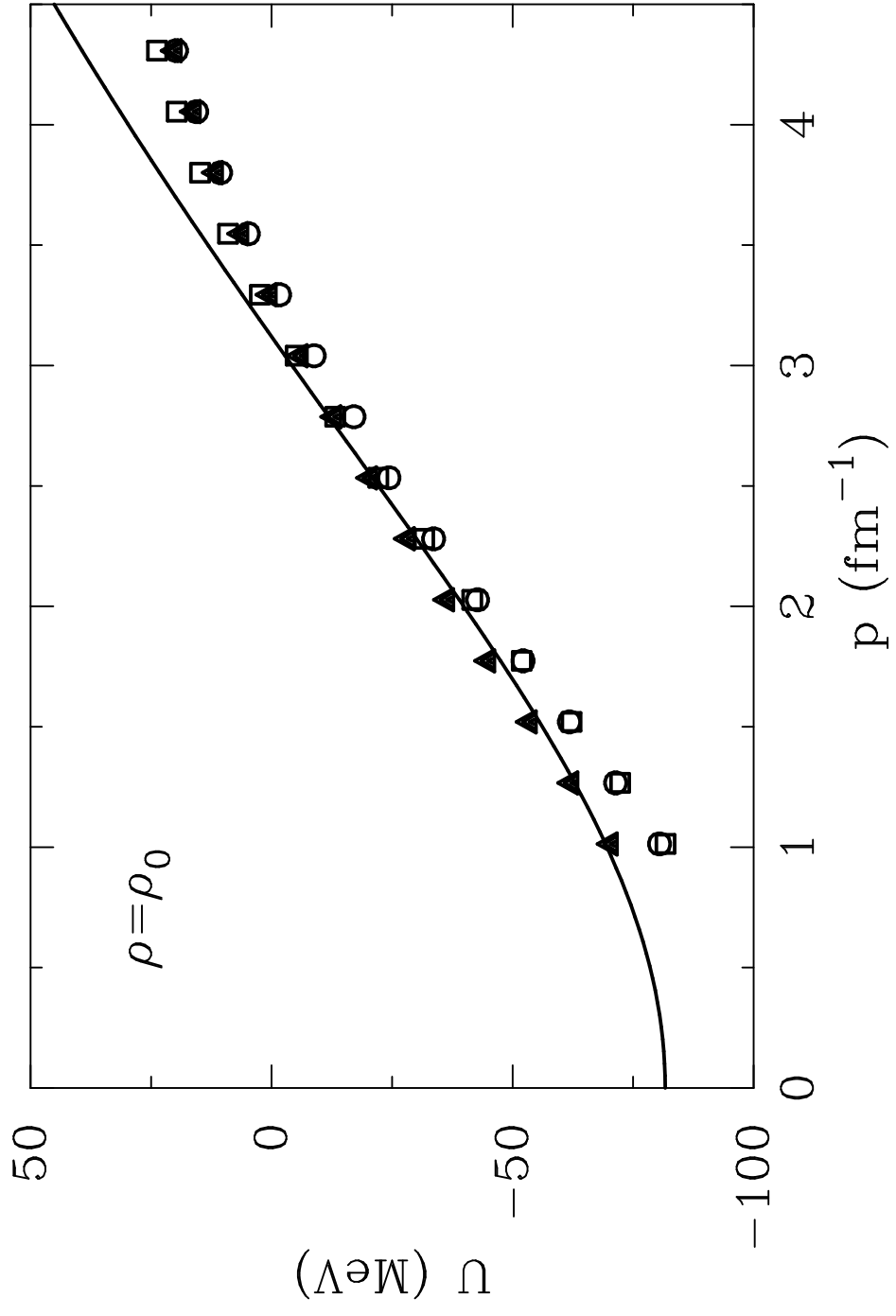
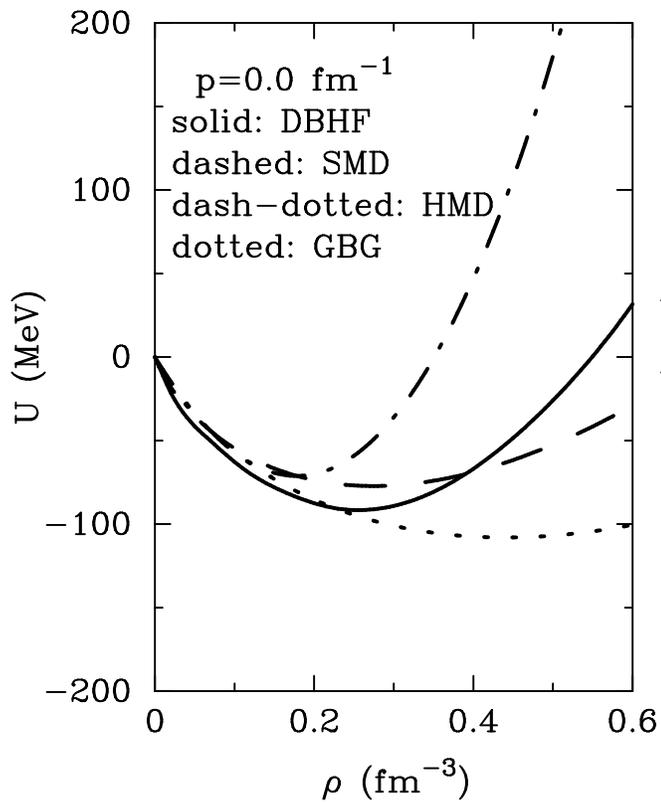
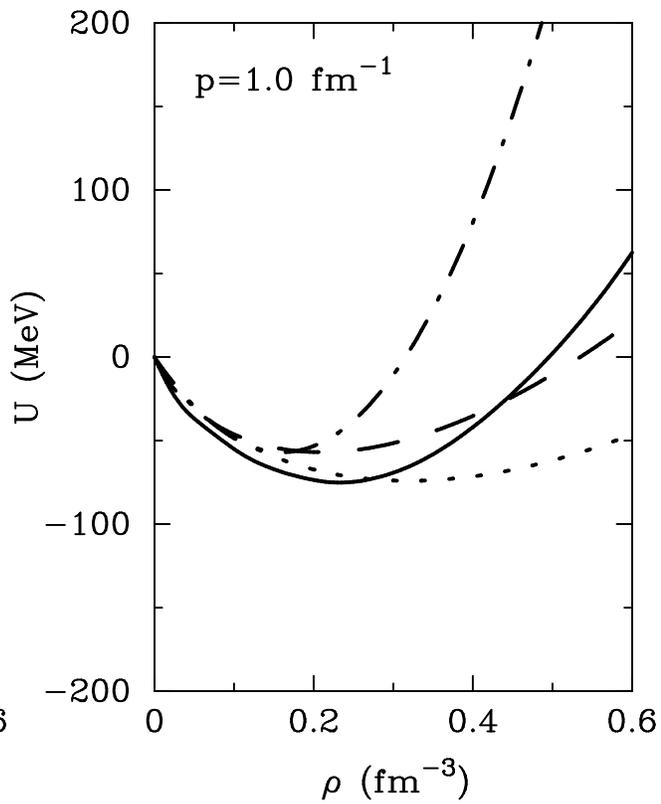
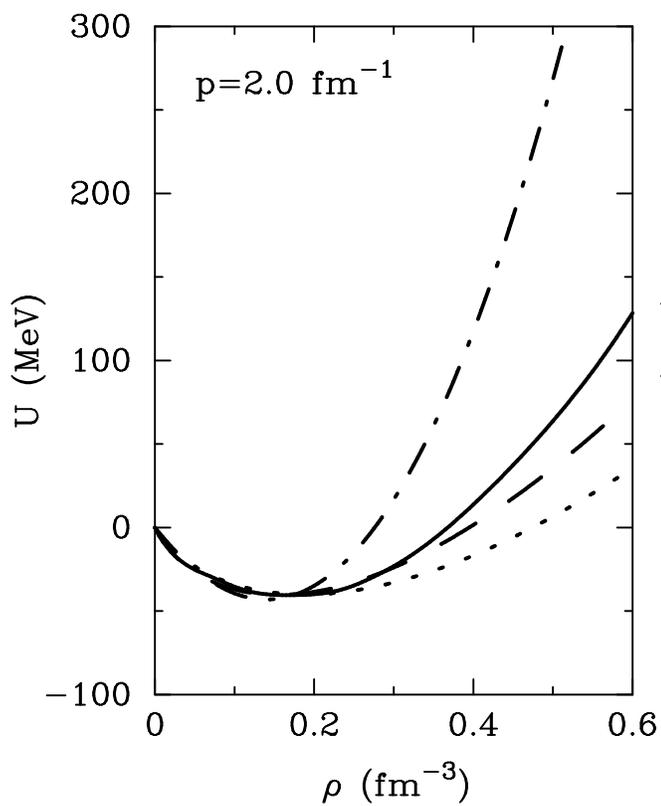
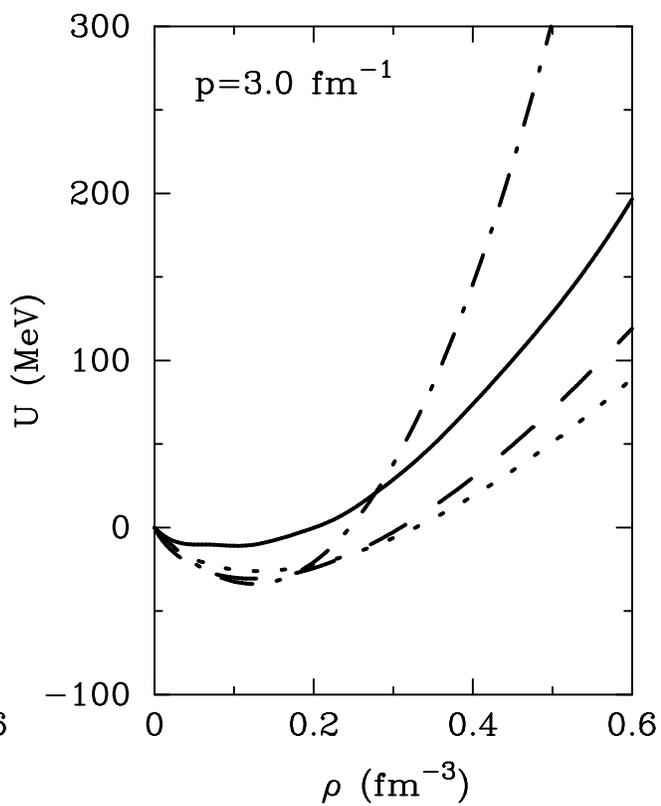
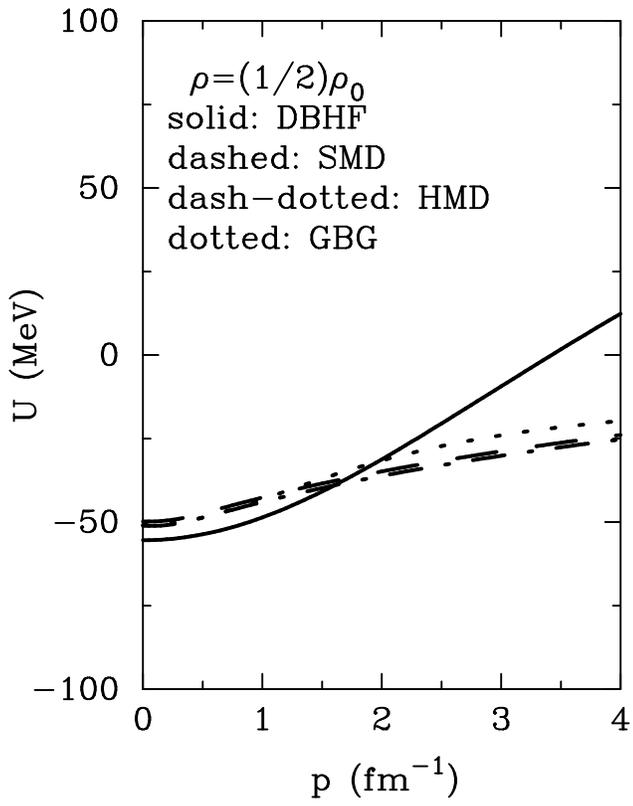
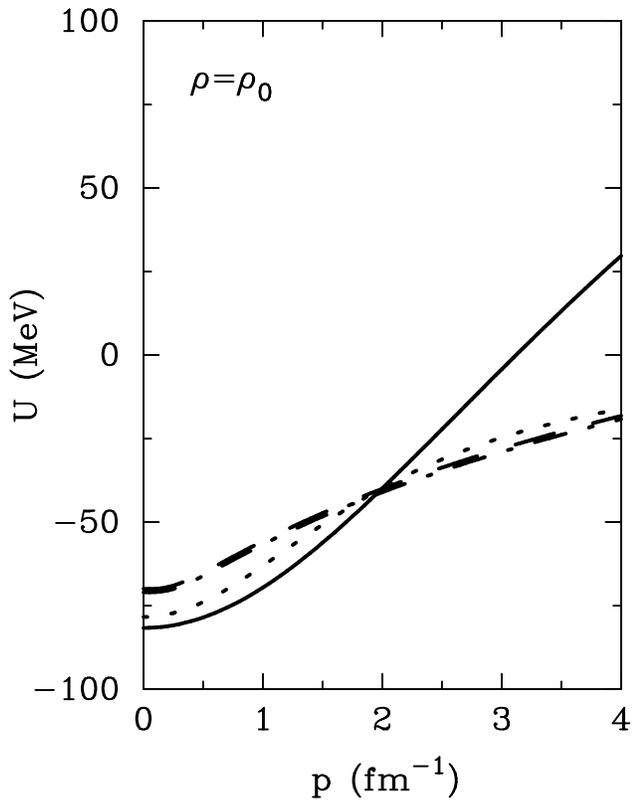
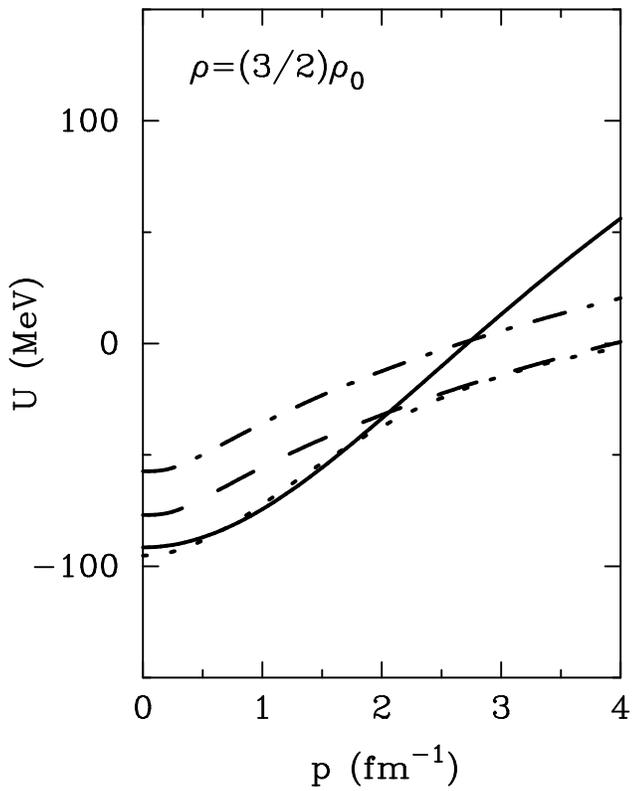
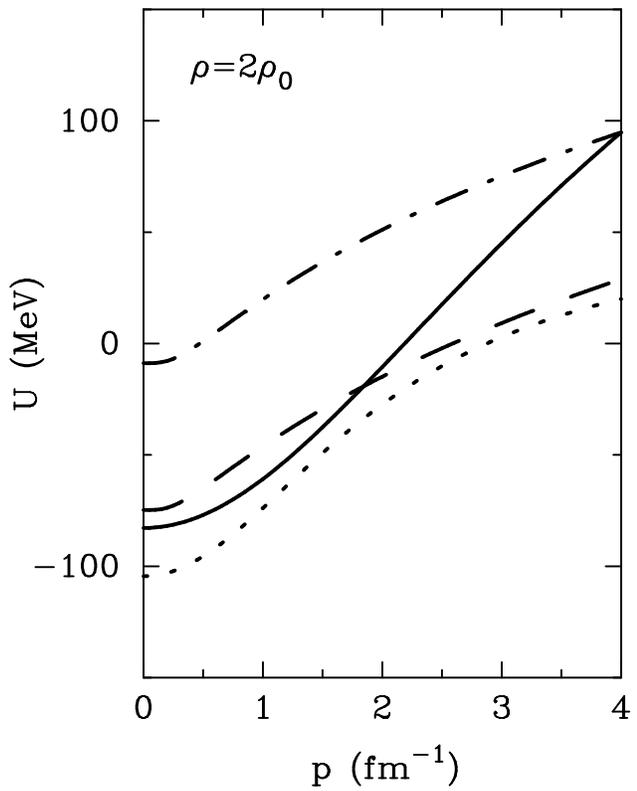
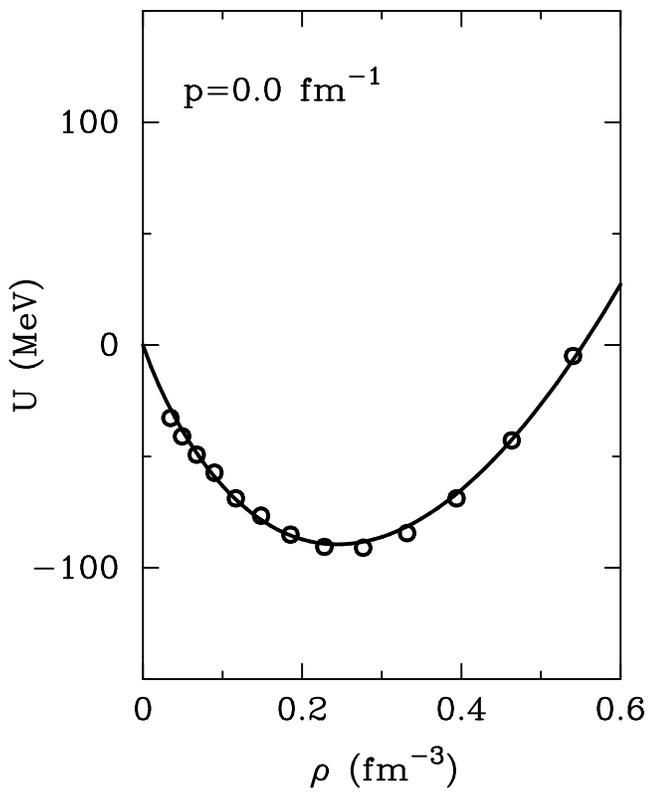
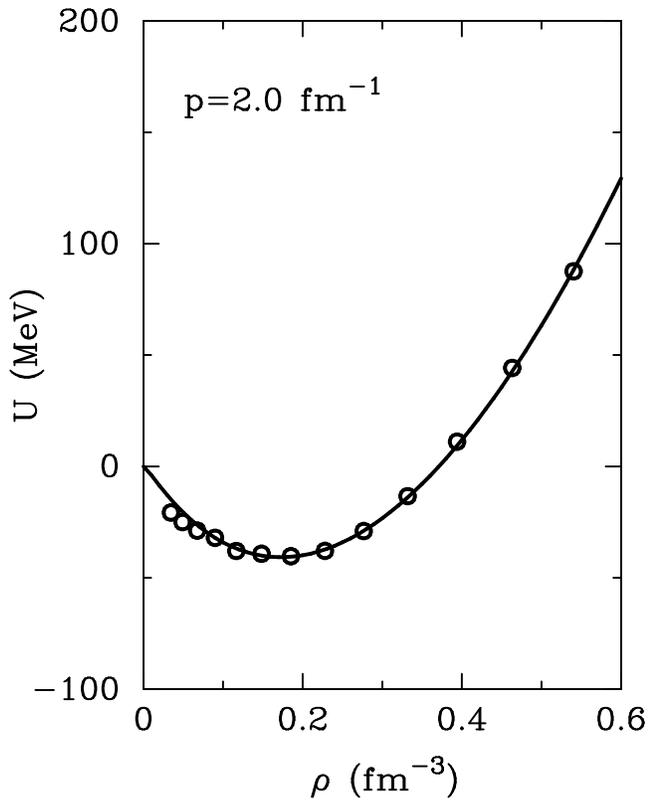
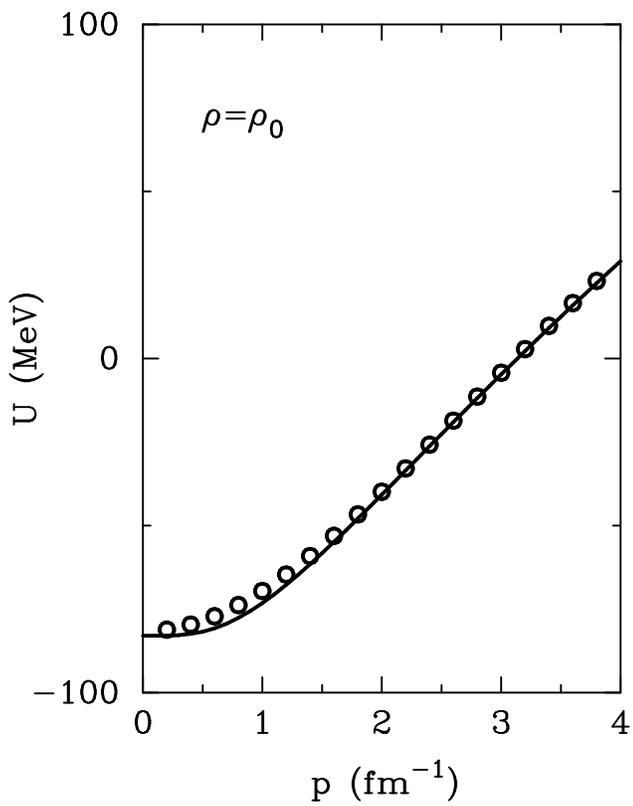
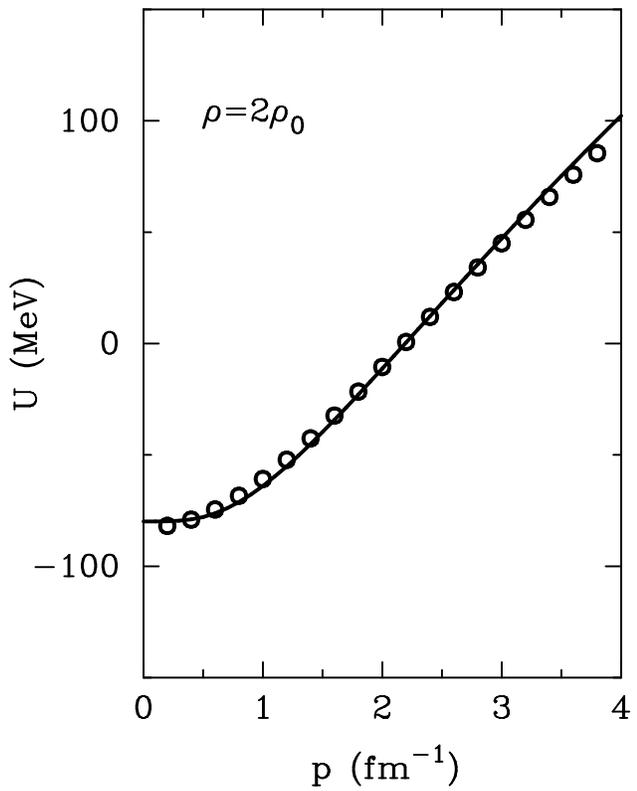
\end{document}